\documentclass{appolb}
\usepackage{epsfig}

\begin{document}
\title{ Diffractive Contribution to $A_\perp$ asymmetry
\thanks{Presented at X International Workshop on Deep
 Inelastic Scattering (DIS2002)\\ Cracow
 30 April - 4 May 2002}%
}
\author{S.V.Goloskokov
\address{Bogoliubov Laboratory of Theoretical  Physics,\\
  Joint Institute for Nuclear Research.
 \\
 Dubna 141980, Moscow region, Russia.}
} \maketitle
\begin{abstract}
We  consider  double spin asymmetries for longitudinally polarized
leptons and transversely polarized protons in diffractive
 $Q \bar Q$ production which is connected with $A_\perp$ asymmetry.
The predicted asymmetry is large and can be used to obtain the
information on the polarized skewed gluon distributions in the
proton.
\end{abstract}
\PACS{PACS 12.38.Bx, 13.60.Hb, 13.88.+e}

\section{Introduction}
Sensitivity of diffractive lepto and photoproduction to the gluon
density in the proton gives an excellent tool to test these
structure functions. Intensive experimental study of diffractive
processes were performed in DESY \cite{zeush1,dijet}. Theoretical
analysis shows that the cross sections of diffractive hadron
production are expressed in terms of skewed parton distributions
(SPD) \cite{rad,ji}. The diffractive charm $Q \bar Q$ production
and $J/\Psi$ production are determined by the gluon SPD ${\cal
F}_\zeta(x)$ because the charm  component in the proton is small.
For light hadron production effects of the quark SPD should be
important for not small $x$. In the polarized case, the
spin-dependent gluon distributions can be investigated. In future,
there will be an excellent possibility of studying spin effects
with transversely polarized target at HERMES and COMPASS.

In this report, we  consider double spin asymmetries for
longitudinally polarized leptons and transversely polarized
protons in diffractive vector $Q \bar Q$ production at high
energies (see \cite{golos02}) which is expressed in terms of the
polarized cross sections
\begin{equation}\label{ds0}
A_\perp =\frac{\left( d \sigma({\rightarrow} {\Downarrow) - d
\sigma({\rightarrow} {\Uparrow}})\right)}{\left( d
\sigma({\rightarrow} {\Downarrow) + d \sigma({\rightarrow}
{\Uparrow}})\right)}.
\end{equation}

At small $x$ the diffractive contribution to the asymmetry
$A_\perp$ determined by the Pomeron exchange should be important.
In the QCD- based models the Pomeron is usually regarded as a
two-gluon state. Within the two-gluon exchange model the two-gluon
coupling with the proton can be written as follows:
\begin{eqnarray}\label{ver}
V_{pgg}^{\alpha\beta}(p,t,x_P,l_\perp)&=& B(t,x_P,l_\perp)
(\gamma^{\alpha} p^{\beta} + \gamma^{\beta} p^{\alpha})\nonumber\\
&+& \frac{i K(t,x_P,l_\perp)}{2 m} (p^{\alpha} \sigma ^{\beta
\gamma} r_{\gamma} +p^{\beta} \sigma ^{\alpha \gamma} r_{\gamma})
+ ... .
\end{eqnarray}
Here the structure proportional to $B(t,...)$ determines the
spin-non-flip contribution. The term $\propto K(t,...)$ leads to
the transverse spin-flip at the vertex. It has been found within
the model approaches \cite{model,golkr} that the ratio $|\tilde
K|/|\tilde B| \sim 0.1$ and has a weak energy dependence (weak $x$
dependence). The weak energy dependence of spin asymmetries in
exclusive reactions is not in contradiction with the experiment
\cite{model}.

\section{Diffractive contribution to $A_\perp$ asymmetry}
The diffractive $Q \bar Q$ production in the lepton-proton
reaction is determined by the photon-two-gluon fusion. The
spin-average and spin-dependent cross section can be written in
the form
\begin{equation}
\label{sigma} \frac{d^5 \sigma(\pm)}{dQ^2 dy dx_p dt dk_\perp^2}=
\left(^{(2-2 y+y^2)} _{\hspace{3mm}(2-y)}\right)
 \frac{C(x_P,Q^2) \; N(\pm)}
{\sqrt{1-4(k_\perp^2+m_q^2)/M_X^2}}.
\end{equation}
Here $C(x_P,Q^2)$ is a normalization coefficient, the functions
$N(\pm)$ are determined by a sum of all graphs integrated over the
gluon momenta $l$ and $l'$. We calculate here the imaginary parts
of the photon-two-gluon fusion amplitudes which are expressed
directly in terms of the functions $B$ and $K$ from (\ref{ver}).

The  function $N(+)$ has the form
\begin{equation}\label{np}
N(+)=\left(|\tilde B|^2+|t|/m^2 |\tilde K|^2 \right)
\Pi^{(+)}(t,k_\perp^2,Q^2).
\end{equation}
 Here
\begin{equation}\label{bqq}
\tilde B \sim \int^{l_\perp^2<k_0^2}_0 \frac{d^2l_\perp
(l_\perp^2+\vec l_\perp \vec r_\perp) }
{(l_\perp^2+\lambda^2)((\vec l_\perp+\vec r_\perp)^2+\lambda^2)}
B(t,l_\perp^2,x_P,...) =  {\cal F}^g_{x_P}(x_P,t,k_0^2),
\end{equation}
and
\begin{equation}\label{kqq}
\tilde K \sim \int^{l_\perp^2<k_0^2}_0 \frac{d^2l_\perp
(l_\perp^2+\vec l_\perp \vec r_\perp) }
{(l_\perp^2+\lambda^2)((\vec l_\perp+\vec r_\perp)^2+\lambda^2)}
K(t,l_\perp^2,x_P,...) =  {\cal K}^g_{x_P}(x_P,t,k_0^2)
\end{equation}
with  $k_0^2 \sim \frac{k_\perp^2+m_q^2}{1-\beta}$. The gluon
structures $\tilde B$($\tilde K$)  are connected with the ${\cal
F}^g_{x_P}(x_P)$(${\cal K}^g_{x_P}(x_P)$) SPD (see
(\ref{bqq},\ref{kqq})). Thus, the functions $B$ and $K$ are the
nonintegrated gluon distributions. The hard part $\Pi^{(+)}$ in
(\ref{np}) can be calculated perturbatively when $k^2_\perp$ is
not small, about $1 \mbox{GeV}^2$ or larger.

The spin-dependent contribution $N(-)$ has two terms proportional
to the scalar products ($\vec k_\perp \vec S_\perp$) and ($\vec Q
\vec S_\perp$) \cite{golos02}. We consider here only the term
proportional to $(\vec Q \vec S_\perp)$ which can be written as
follows:
\begin{equation}\label{nm}
N(-)=\sqrt{\frac{|t|}{m^2}} \left(\tilde B \tilde K^*+\tilde B^*
\tilde K\right) \left[ \frac{(\vec Q \vec S_\perp)}{m}
\Pi^{(-)}_Q(t,k_\perp^2,Q^2) \right].
\end{equation}
The other term $\propto (\vec k_\perp \vec S_\perp)$ has been
discussed in \cite{golos02}.

\bigskip
\begin{figure}[h]
\epsfxsize=10cm \centerline{\epsfbox{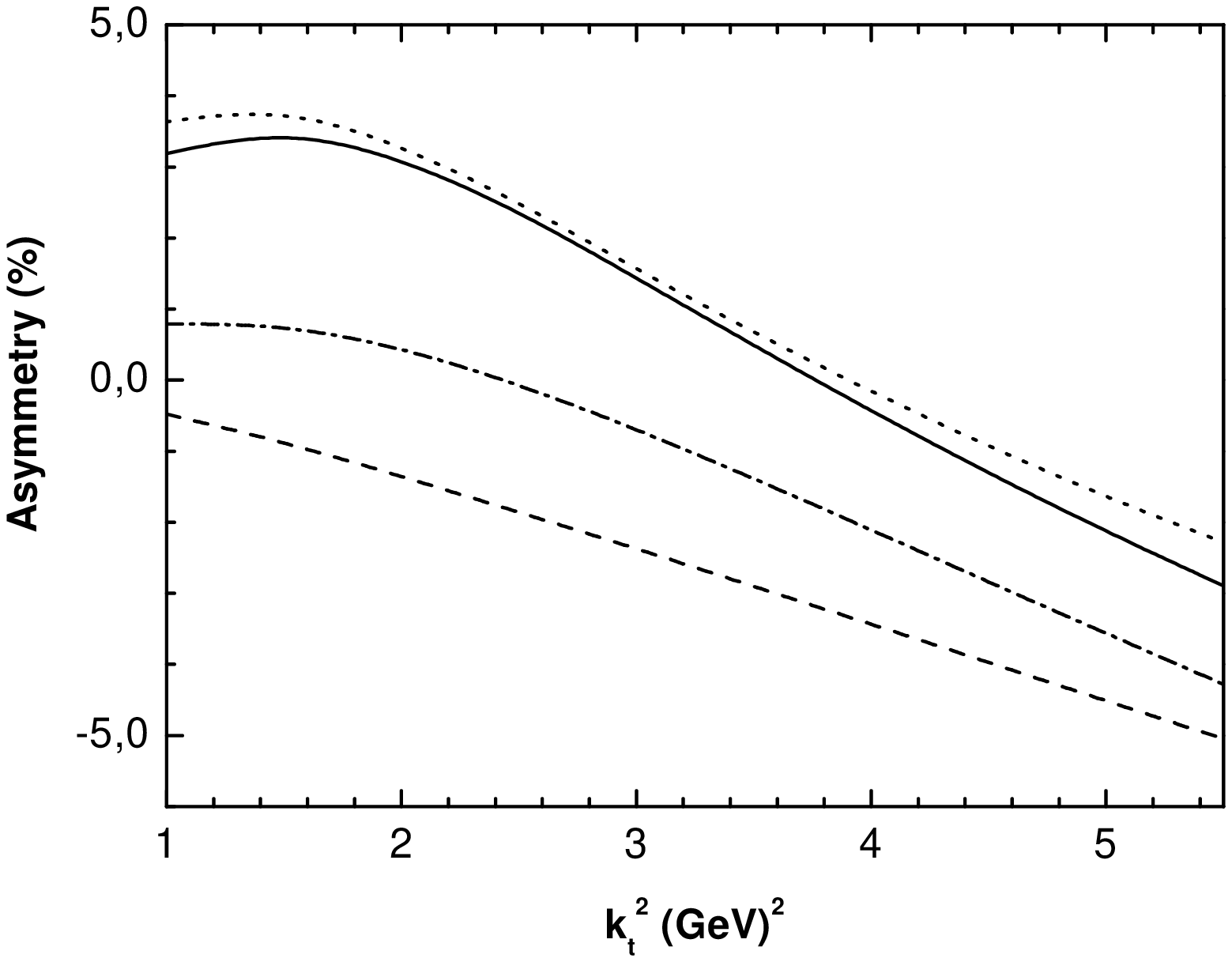}}
\caption{$A_\perp$ asymmetry in diffractive light $Q \bar Q$
production at $\sqrt{s}=20 \mbox{GeV}$ for $x_P=0.1$, $y=0.5$,
$|t|=0.3 \mbox{GeV}^2$: dotted line-for $Q^2=0.5 \mbox{GeV}^2$;
solid line-for $Q^2=1 \mbox{GeV}^2$; dot-dashed line-for $Q^2=5
\mbox{GeV}^2$; dashed line-for $Q^2=10 \mbox{GeV}^2$.}
\end{figure}
The asymmetry is determined by the ratio
$A_\perp=\sigma(-)/\sigma(+)$. At small $x$ the gluon structure
functions have large imaginary part. In this case the asymmetry
can be approximated as
\begin{equation}\label{cltqq}
 A_\perp \sim C_\perp \frac{\tilde K}{\tilde B}= C_\perp \frac{{\cal K}^g_\zeta(\zeta)}
 {{\cal F}^g_\zeta(\zeta)}
 \;\;\;\mbox{with} \;\zeta=x_P
\end{equation}

In numerical calculations we use a simple parameterization of the
SPD as a product of the form factor and the ordinary gluon
distribution \cite{golos02}. In our estimations we use the value
$|\tilde K|/|\tilde B| \sim 0.1$. We analyze the case when the
$A_{\perp}$ asymmetry  has a maximal value (the momentum $\vec
Q_\perp$ is parallel to the target polarization $\vec S_\perp$).
The predicted $A_\perp$ asymmetry in diffractive light $Q \bar Q$
production at $\sqrt{s} =20 \mbox{GeV}$ is shown in Fig. 1. This
asymmetry is not small for $Q^2 \sim (0.5-1) \mbox{GeV}^2$. The
$A_{\perp}$ asymmetry has a strong mass dependence. For heavy
quark production this asymmetry becomes negative, Fig. 2.

\bigskip
\begin{figure}[h]
\epsfxsize=10cm \centerline{\epsfbox{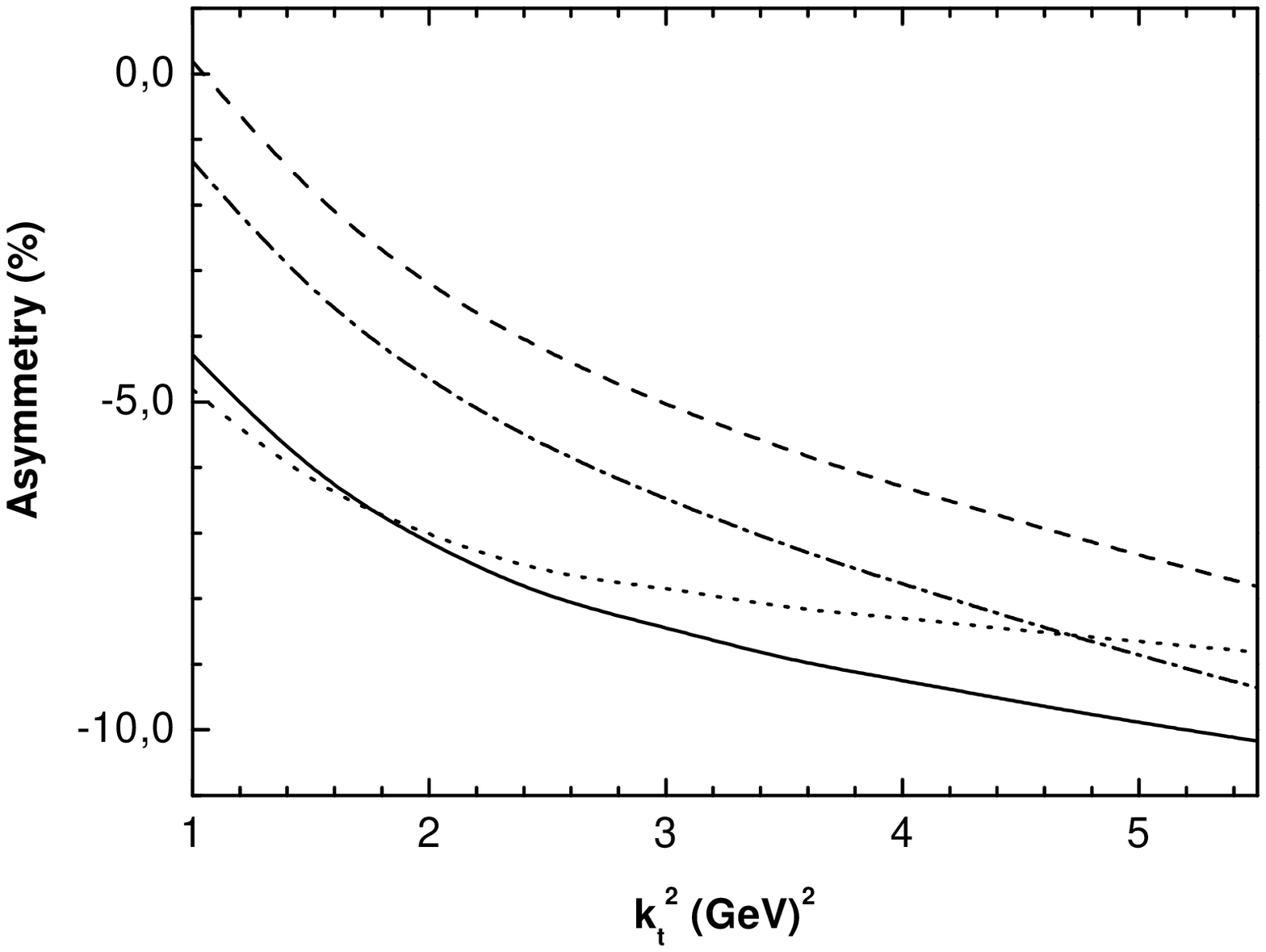}}
\caption{$A_\perp$ asymmetry in diffractive heavy quark
production. Lines are the same as in Fig.1}
\end{figure}

It is interesting to have predictions for a light quark production
at the SLAC and HERMES low--energy range $\sqrt{s} \sim 7
\mbox{GeV}$. Here the "window" for perturbative calculation is
quite small. Really, it can be found in this case that the maximum
value of the transverse momentum is limited by $k^2_\perp <2
\mbox{GeV}^2$ because we have the restriction  $k^2\le M_X^2/4$
from (\ref{sigma}). In fixed--target experiments, it is usually
difficult to detect the final hadron and determine the momentum
transfer. In this case, we estimate the asymmetry integrated over
momentum transfer
\begin{equation}\label{intasy}
\bar A_\perp=
\frac{\int_{t_{min}}^{t_{max}}\,\sigma(-)\,dt}
{\int_{t_{min}}^{t_{max}}\,\sigma(+)\,dt}
.
\end{equation}
$t_{min} \sim 0$ ; $t_{max}=4 \mbox{GeV}^2$. The predicted
integrated asymmetry  is  about 3\%. Note that we have calculated
here only the gluon contribution to the asymmetry. At HERMES
energies the contribution of the quark SPD to the $A_\perp$
asymmetry should be important.

\section{Conclusion}
To conclude, we would like to emphasize that the diffraction
contribution to the $A_\perp$ asymmetry is found to be
proportional to the ratio of ${\cal K}/{\cal F}$ structure
functions. The predicted coefficient $C_\perp$ in (\ref{cltqq}) is
not small, about 0.3-0.5. This shows the possibility  of studying
the transverse distribution ${\cal K}^g_{x_P}(x_P,t)$ in future
experiments with a transversely polarized target (HERMES, COMPASS
and future eRHIC facilities). These results could be applicable to
 reactions with heavy quarks. For processes with light hadron production,
our predictions can be used in the small $x$ region ($x \le 0.1$
e.g.) where the contribution of the quark SPD is expected to be
small. The recoil particle detector is needed to distinguish the
diffractive events. Really, in the case when the recoil detector
is absent, the diffractive events are detected together with
nondiffractive ones. The measured asymmetry in this case looks
like
\begin{equation}\label{as}
A_{exp}=\frac{\Delta \sigma_{ND}+\Delta \sigma_{D}}{\sigma_{ND}+
\sigma_{D}}=A_{ND} (1-R) + A_D R,\;\;
R=\frac{\sigma_{D}}{\sigma_{ND}+ \sigma_{D}}.
\end{equation}
Here $A_{ND}=\Delta \sigma_{ND}/\sigma_{ND}$ and $A_D=\Delta
\sigma_{D}/\sigma_{D}$. The ratio  $R$ integrated over $x$  has
been found at HERA to be about 0.20--0.30 \cite{diff}. In this
case, the diffractive contribution to asymmetry will be smaller by
the factor 3-5.

This work was supported in part by the Russian Foundation for
Basic Research, Grant 00-02-16696 and by the Bogoliubov-Infeld
programme.

\end{document}